\documentclass[11pt]{article}             
\usepackage{osid}                         %

\usepackage{graphics}
\usepackage{graphicx}
\usepackage{bm}
\usepackage{latexsym}
\usepackage{color}
\usepackage{mathrsfs}
\usepackage{braket}
\usepackage{enumerate}
\usepackage[usenames,dvipsnames]{xcolor}
\usepackage[colorlinks=true,citecolor=blue,linkcolor=blue,urlcolor=blue,backref=true]{hyperref}
\usepackage{float}
\usepackage[raggedright,tight]{subfigure}  
\usepackage{amsmath}
\usepackage{amsfonts}
\usepackage{braket}
\usepackage[normalem]{ulem}
\usepackage{cancel}
\usepackage{mathtools}
\usepackage{appendix}
\usepackage{bm}

\title{The gauge-relativity of quantum light, matter, and information}
\author{Adam Stokes \\{\footnotesize\it School of Mathematics, Statistics, and Physics, Newcastle University, Newcastle upon Tyne NE1 7RU, United Kingdom \& adamstokes8@gmail.com}\\[2ex]
Hannah Riley
                     \\{\footnotesize\it Department of Physics and Astronomy, University of Manchester, Oxford Road, Manchester M13 9PL, United Kingdom \& hannah.riley-6@postgrad.manchester.ac.uk}\\[2ex]
        Ahsan Nazir\\{\footnotesize\it Department of Physics and Astronomy, University of Manchester, Oxford Road, Manchester M13 9PL, United Kingdom \& ahsan.nazir@manchester.ac.uk}
         }

\begin{document}

\maketitle
\begin{abstract}
We describe the physical relativity of light and matter quantum subsystems, their correlations, and energy exchanges. We examine the most commonly adopted definitions of atoms and photons, noting the significant difference in their localisation properties when expressed in terms of primitive manifestly gauge-invariant and local fields. As a result, different behaviours for entanglement generation and energy exchange occur for different definitions. We explore such differences in detail using toy models of a single photonic mode interacting with one and two dipoles.  
\end{abstract}

\noindent {\em Dedication}: Two of us gave lectures in Stockholm in November 2022, about a week before G\"oran Lindblad's
passing away. We would like to think that he would have found the topics we address interesting,
and we regret never having had the opportunity to discuss them with him. We dedicate this contribution to
Lindblad's memory.\\

\section{Introduction}

Elementary microscopic systems possess wavelike nature so their physical states are represented using vectors, which can be superposed. A bipartite quantum system possesses superpositions that are not separable into pure states of the individual parts. These entangled states are one of quantum theory's most notable distinguishing features, but for the same reason that such states exist, the quantum systems between which entanglement occurs are inherently {\em relative}. 

In a theory built using a linear (vector) space, relativity occurs whenever the meaning assigned to elements of the space depends on the chosen basis (frame). In different frames a quantum subsystem of a bipartite composite generally constitutes distinct sets of physical states and observables \cite{stokes_implications_2022}. The most relevant quantum subsystems are determined by the set of operationally available interactions and measurements \cite{stokes_implications_2022,zanardi_quantum_2004,viola_entanglement_2007,zanardi_virtual_2001,hoehn_quantum_2021,harshman_observables_2011,harshman_tensor_2007,cai_entanglement_2020}.

In quantum electrodynamics (QED), gauge changes are implemented by unitary transformations between frames that do not preserve the state space partition into subsystems. The ensuing physical relativity of quantum light and matter can only be ignored in the calculation of (non-dynamical) scattering matrix ($S$-matrix) elements and in certain dynamical predictions that, within standard weak-coupling and Markovian regimes, are found to depend solely on $S$-matrix elements. An example is the Lindblad master equation for a dipole in a photonic environment \cite{stokes_implications_2022}.

In all other situations, quantum subsystem (gauge)-relativity is a fundamental feature that cannot be ignored. In particular, both atom-photon and atom-atom entanglement are, by definition, gauge relative. This aspect of QED will be the focus of the present study. In Sec.~\ref{relofss}, we discuss the physical differences between quantum subsystems defined relative to different gauges. We find that the quantum information-theoretic notion of locality based on tensor-product structure is generally distinct to the field-theoretic notion in which local properties in spacetime are assigned using physical fields. As such, distinct definitions of an atom as a quantum subsystem generally differ markedly in their field-theoretic spacetime localisation properties. In Sec.~\ref{toy} we discuss the implications of this for interatomic entanglement, with relevance to the electromagnetic analog of gravitationally induced entanglement between two masses. The latter has been proposed as a witness for the quantisation of gravity and the existence of an associated mediating massless field (gravitons) \cite{marletto_gravitationally_2017,bose_spin_2017,fragkos_inference_2022}.

In order to begin to explore the significance of subsystem gauge-relativity for interatomic entanglement, information, and energy exchanges, we study simple toy models consisting of a single-mode interacting with one and two dipoles. We compare entropies and populations for atomic and photonic subsystems defined relative to a gauge that varies continuously with a single real parameter. While fundamentally gauge-invariant, subsystems properties, including entanglement, are found to be strongly gauge-relative for sufficiently large coupling strengths. The different underlying mechanisms leading to entanglement between different physical subsystems are contrasted. We summarise our findings in Sec.~\ref{conc} Throughout we use natural units with $\hbar=c=\epsilon_0=1$.

\section{Relativity of quantum systems}\label{relofss}

\subsection{Composite quantum systems and Hilbert space frames}

To understand the relativity of quantum subsystems an analogy with special relativity proves most useful. Any two inertial frames $A$ and $B$ in spacetime $M$ are connected by a Lorentz transformation $\Lambda$. 
For two events $Y$ and $Z$ with coordinate representations $y$ and $z$ in $M$, the Minkowski inner-product $g$ satisfies $g(y,z)=g(\Lambda y,\Lambda z)$. Yet, the spatial (temporal) intervals between $Y$ and $Z$ found using frames $A$ and $B$ are generally distinct; $\Delta x_A \neq \Delta x_B$ ($\Delta t_A \neq \Delta t_B$). The intervals $\Delta x_A$ and $\Delta t_A$ ($\Delta x_B$ and $\Delta t_B$) are respectively measured using a ruler and a clock at rest in frame $A$ ($B$). Each frame yields a set of unique and meaningful physical predictions, but the two sets of predictions refer to different experiments. It is evident that, operationally, the partition $M=T\oplus S$ of spacetime into ``space" $S$ and ``time" $T$ can only be defined relative to an inertial frame.

In quantum theory the pure states of a physical system are represented by vectors in a Hilbert space ${\cal H}$ and the observables are represented by operators in the algebra ${\cal A}=\{O:{\cal H}\to {\cal H} ~|~ O=O^\dagger\}$. Mixed states are represented by operators in the subspace ${\cal D}$ of positive Hermitian operators with unit trace. All physical predictions are obtained using the Hilbert space inner-product, and a unitary transformation $U$ between two frames (orthonormal bases) $A$ and $B$ in ${\cal H}$ leaves the inner-product between any two vectors unchanged; $\braket{\psi|\phi}=\braket{U\psi | U\phi}$. 

The question arises as to whether, in quantum theory, there might occur a division of a Hilbert space into parts that is not left invariant by certain unitary transformations, such that the physical meanings of the individual parts become inherently relativised. The answer is affirmative. The partition of a composite quantum system into subsystems, which is the basic means by which we understand energy and information exchanges, is of precisely this nature.

Composite quantum systems are constructed using the tensor product, because it extends the inner-product in a way that is consistent with the probabilistic postulates of quantum theory (Born rule). Probabilities associated with independent subsystems, $1$ and $2$, are those associated with independent events, that is, $(\bra{\psi_1}\otimes \bra{\psi_2})(\ket{\phi_1}\otimes \ket{\phi_2}) = \braket{\psi_1|\phi_1}\braket{\psi_2|\phi_2}$. If $({\cal H}_i,{\cal D}_i,{\cal A}_i),~i=1,2$ denotes a pair of quantum systems, then a composite system can be constructed with Hilbert space ${\cal H}={\cal H}_1\otimes {\cal H}_2$, density operator space ${\cal D}={\cal D}_1\otimes {\cal D}_2$, and observable algebra ${\cal A}={\cal A}_1\otimes {\cal A}_2$. Conversely, given a Hilbert space ${\cal H}$ that can be divided as ${\cal H}={\cal H}_1\otimes {\cal H}_2$, each ${\cal H}_i$ defines a quantum (sub)system. An observable of subsystem $1$ represented by an operator $O\in {\cal A}_1$ is represented within ${\cal A}$ by the operator $O\otimes I_2$ where $I_2$ is the identity in ${\cal A}_2$. If $\rho\in {\cal D}$ represents the state of the composite then $\rho_1 ={\rm tr}_2\rho \in {\cal D}_1$ can be said to represent the state of subsystem $1$, because it suffices to compute the physical predictions for any observable of subsystem $1$, that is, ${\rm tr}(O\otimes I_2\rho) = {\rm tr}(O\rho_1)$.

Now consider a unitary transformation of a composite quantum system, $U:{\cal H}\to {\cal H}$. If we begin by representing a physical state ${\cal S}$ and observable ${\cal O}$ by a density operator $\rho \in {\cal D}$ and Hermitian operator ${O} \in {\cal A}$ respectively, then we immediately discover that it must be equally valid to represent the same state and observable by the operators $\rho'=U\rho U^\dagger$ and $O'=UOU^\dagger$. This is because the physical prediction $\langle {\cal O} \rangle_{\cal S}$ for the average value of ${\cal O}$ in the state ${\cal S}$ can be found using either pair, that is,
\begin{align}\label{gi}
{\rm tr}(O\rho)=\langle {\cal O} \rangle_{\cal S} = {\rm tr}(O'\rho').
\end{align}
If we label the frames connected by $U$ by $A$ and $B$, then we can conclude that $\rho$ ($\rho'$) and $O$ ($O'$) are the representations of ${\cal S}$ and ${\cal O}$ with respect to the frame $A$ ($B$). Let us suppose however, that $U$ does not possess the form $U=U_1\otimes U_2$, then although $\rho$ is unitarily equivalent to $\rho'$ and so both of these operators represent the same physical state, it is not the case that $\rho_1={\rm tr}_2\rho$ is unitarily equivalent to $\rho_1'={\rm tr}_2\rho'$. Therefore $\rho_1$ and $\rho_1'$ must, in general, represent different physical states. Similarly, if in frame $A$ an observable ${\cal O}$ is represented by an operator of the form $O\otimes I_2$ belonging to the quantum subsystem $1$, then in frame $B$ it is represented by the operator $O'=UO_1\otimes I_2 U^\dagger \neq O'\otimes I_2$, which does not belong to the quantum subsystem $1$. We conclude that quantum subsystems are physically {\em relative}. The quantum system $({\cal H}_i,{\cal D}_i,{\cal A}_i)$ must represent a different collection of physical states and observables in any two frames $A$ and $B$ of ${\cal H}$ that are connected by a transformation $U\neq U_1\otimes U_2$.

In the same way that the parts $T$ and $S$ of $M$ cannot be assigned physical (operational) meaning independent of the considered inertial frame in $M$, the parts ${\cal H}_1$ and ${\cal H}_2$ cannot be assigned physical (operational) meaning independent of the considered frame in ${\cal H}$. We may continue to speak of the quantum subsystems $1$ and $2$ in every frame of ${\cal H}$, in the same way that we continue to use the same labels ``space" and ``time" in every inertial frame of $M$. But in the same way that measurement of ``space" refers to two different operational procedures in frames $A$ and $B$ (holding a ruler at rest in frame $A$ versus holding a ruler at rest in frame $B$) we must recognise that the label $1$ generally refers to different states and observables and thereby different operational procedures in frames $A$ and $B$ of ${\cal H}$.

\subsection{The relativity of atoms and photons}\label{relatphot}

\subsubsection{Gauss' law and electric field decompositions}

The importance of quantum subsystem relativity is especially evident in quantum electrodynamics, which is a constrained theory that already possesses a fundamental mathematical redundancy in the form of gauge-freedom. Gauss' law, $\nabla \cdot {\bf E} =\rho$, stating that material electric charges with density $\rho$ are sources (or sinks) of an electric field ${\bf E}$, is a non-dynamical constraint. It makes a complete separation of material and electromagnetic degrees of freedom into distinct subsystems impossible, because it implies that some part of the electric field must be material.

Recall that every $3$-vector field ${\bf V}$ that decays sufficiently fast at infinity admits a unique (Helmholtz) decomposition, ${\bf V}={\bf V}_{\rm T}+{\bf V}_{\rm L}$, into a transverse part (such that $\nabla \cdot {\bf V}_{\rm T}=0$) and a longitudinal part (such that $\nabla \times {\bf V}_{\rm L}={\bf 0}$). Adding any transverse field to a particular solution of Gauss' law, yields a second solution. The most general decomposition of the electric field is therefore ${\bf E}=-{\bf \Pi}-{\bf P}$ where $-\nabla \cdot {\bf P}=\rho$ and ${\bf \Pi}=-{\bf E}_{\rm T}-{\bf P}_{\rm T}$. The Helmholtz decomposition of ${\bf E}$ is a special case obtained by choosing ${\bf P}_{\rm T}={\bf 0}$, such that ${\bf P}=-{\bf E}_{\rm L} = -\nabla (\nabla^{-2}\rho)$. More generally, any ${\bf P}$ that depends only on material charge and current densities can be said to be the material part of the electric field, of which ${\bf P}_{\rm L}$ is merely one (highly non-local) possible definition. As will be seen in what follows, the remaining transverse component ${\bf \Pi}$ is photonic.

\subsubsection{Lorenz gauge}

Writing the electric and magnetic fields as ${\bf E}=-\partial_t {\bf A} - \nabla A_0$ and ${\bf B}=\nabla \times {\bf A}$, where $A_0$ and ${\bf A}$ are scalar and vector potentials, implies that the homogeneous Maxwell equations are automatically satisfied, but under a gauge transformation $A_\mu \to A_\mu-\partial_\mu \chi$ the fields are invariant. The redundancy within the formalism is eliminated by fixing a gauge, that is, by fixing each potential component as a specified function of physical degrees of freedom that cannot be freely chosen. A common choice in high-energy physics is the Lorenz gauge, which maintains manifest covariance and is defined classically by the condition $L=\partial_\mu A^\mu =0$. At least ostensively, this gauge retains complete separation between material and electromagnetic degrees of freedom. Scalar photons (associated with $A_0$), longitudinal photons (associated with ${\bf A}_{\rm L}$ and ${\bf E}_{\rm L}$), transverse photons (associated with ${\bf A}_{\rm T}=(\nabla\times )^{-1}{\bf B}$ and ${\bf E}_{\rm T}$), and matter, each define a separate quantum subsystem. Separation is not truly achieved however, because the theory is constrained.

The classical Lorenz gauge condition $L=0$ implies $\square A_0 =\rho$ where $\square = \partial_t^2-\nabla^2$. When substituted into $G=-\nabla\cdot {\bf E}+\rho$ one obtains $G=\partial_t L$, showing that Gauss' law is satisfied if and only if the Lorenz gauge condition is stable in time. Scalar and longitudinal electromagnetic degrees of freedom cannot be truly independent of matter. In the quantum theory, the Lorenz gauge choice must define a subsidiary condition that the vectors $\ket{\psi}$ representing physical states have to satisfy and that is stable in time, namely $[a_{\rm L}({\bf k})-a_s({\bf k})-\lambda({\bf k})]\ket{\psi}=0$ \cite{cohen-tannoudji_photons_1989}. The operators $a_{\rm L}({\bf k})$ and $a_s({\bf k})$ annihilate longitudinal and scalar photons (of momentum ${\bf k}$) respectively, and $\lambda({\bf k})=-{\tilde \rho}\left({\bf k})/(\omega\sqrt{2\omega}\right)$ where $\omega=|{\bf k}|$ and a tilde denotes the Fourier transform. It follows that in the Lorenz gauge a state represented by a vector $\ket{0_{\rm L}}\otimes \ket{0_s}\otimes \ket{\psi_m}$ with no scalar or longitudinal photons and an arbitrary material part $\ket{\psi_m}$, is not physical. The subsystem we have called ``matter" in this gauge cannot be separated from the subsystems that we have called scalar and longitudinal photons.

Imposing the Lorenz subsidiary constraint within the average electric field in a physical state yields the Helmholtz decomposition $\langle{\bf E}\rangle=\langle{\bf E}_{\rm T}+{\bf E}_{\rm L}\rangle$ where ${\bf E}_{\rm L}:=\nabla (\nabla^{-2}\rho)$ and ${\bf E}_{\rm T}:=-{\dot {\bf A}}_{\rm T}$. As will be seen below, this decomposition is automatically obtained at the operator level within the Coulomb gauge defined by $\nabla \cdot {\bf A}=0$. Transformation to the Coulomb gauge is implemented using a unitary operator $T$, which displaces $a_s({\bf k})$ as $Ta_s({\bf k})T^{-1}=a_s({\bf k})+\lambda({\bf k})$, but leaves $a_{\rm L}({\bf k})$ and $\lambda({\bf k})$ unchanged \cite{cohen-tannoudji_photons_1989}. A physical state represented by $\ket{\psi}$ in the Lorenz gauge is represented by $T\ket{\psi}=\ket{\psi'}$ in the Coulomb gauge, and it must satisfy $T[a_{\rm L}({\bf k})-a_s({\bf k})-\lambda({\bf k})]\ket{\psi}=[a_{\rm L}({\bf k})-a_s({\bf k})]\ket{\psi'}=0$. In the Coulomb gauge, restricting ones attention to physical states amounts to ignoring scalar and longitudinal photons completely \cite{cohen-tannoudji_photons_1989}. 

The gauge-fixing transformation $T$ is not of product form, but it leaves the Hilbert space bipartition into photons of the transverse electric field and the corresponding definition of physical matter unchanged. In the Coulomb gauge this matter constitutes a single unconstrained physical system whereas in the Lorenz gauge it constitutes a constrained tripartite system. The two gauges are strictly equivalent and while no non-relativistic approximation of the Maxwell field exists, any gauge can be considered with either relativistic or non-relativistic matter.

\subsubsection{Coulomb gauge}

Separating the longitudinal electric field from the transverse field and defining the former as a material operator is advantageous if one seeks to describe stable bound systems of charges, such as atoms and molecules. The Coulomb energy 
\begin{align}
V^{\rm Coul}={1\over 2}\int d^3x {\bf E}_{\rm L}({\bf x})^2 = {1\over 2}\int d^3 x \int d^3 x' {\rho({\bf x})\rho({\bf x}')\over 4\pi|{\bf x}-{\bf x}'|}
\end{align}
offers an excellent approximation of the retarded near-field interaction that is dominant in binding opposite charges together at atomic length scales. It can be used to define a material energy operator that possesses discrete eigenvalues and eigenvectors. The remaining electromagnetic interactions are treated as a perturbation that results in transitions between material energy levels when transverse photons are emitted and absorbed. This basic idea is the fundamental means by which we understand quantum light-matter physics.

We will make a non-relativistic approximation of the motion of charges for simplicity, but all results carry over to the case of possibly relativistic charge motion. We consider a single atom consisting of a charge $q$ at a point ${\bf r}$ bound to a charge $-q$ stationary at the origin. The charge and current densities are $\rho({\bf x})=q\delta({\bf x}-{\bf r})-q\delta({\bf x})$ and ${\bf J}({\bf x})=q[{\dot {\bf r}}\delta({\bf x}-{\bf r}) +\delta({\bf x}-{\bf r}){\dot {\bf r}}]/2$. Ignoring infinite Coulomb self-energies the Coulomb energy is $V^{\rm Coul}=V({\bf r})=-q^2/(4\pi r)$. The energy of the atom is $H_m = {\bf p}^2/(2m)+V({\bf r})$ where ${\bf p}$ is canonically conjugate to ${\bf r}$, i.e., $[r_i,p_j]=i\delta_{ij}$. The canonical operators ${\bf r}$ and ${\bf p}$, and the energy $H_m$ belong to a material algebra ${\cal A}_m$. The eigenvalue equation $H_m\ket{n} =\omega_n \ket{n}$ defines vectors $\{\ket{n}\}$ that span a Hilbert space ${\cal H}_m$ and that can also be used to write down arbitrary density operators comprising a space ${\cal D}_m$. 

Photons are defined using the gauge-invariant transverse vector potential ${\bf A}_{\rm T}$ and the canonically conjugate momentum ${\bf \Pi}$ by 
\begin{align}\label{phot}
a_\lambda({\bf k}) = {{\bf e}_\lambda ({\bf k})\over \sqrt{2\omega}}\cdot \left[\omega {\tilde {\bf A}}_{\rm T}({\bf k})+i{\tilde {\bf \Pi}}({\bf k})\right].
\end{align}
The ${\bf e}_\lambda ({\bf k}),~\lambda=1,2$ are mutually orthogonal unit vectors orthogonal to ${\bf k}$ and a tilde is used to denote the Fourier transform. The canonical operators belong in an algebra ${\cal A}_{\rm ph}$ and satisfy $[A_{{\rm T},i}({\bf x}),\Pi_j({\bf x}')]=i\delta_{ij}^{\rm T}({\bf x}-{\bf x}')$ while the photonic operators satisfy $[a_\lambda({\bf k}),a^\dagger_{\lambda'}({\bf k}')]=\delta_{\lambda\lambda'}\delta({\bf k}-{\bf k}')$. The photonic energy is
\begin{align}
H_{\rm ph}&={1\over 2}\int d^3 x\left[{\bf \Pi}({\bf x})^2+{\bf B}({\bf x})^2\right] = \int d^3k \sum_\lambda\omega\left(a^\dagger_\lambda({\bf k}) a_\lambda({\bf k})+{1\over 2}\right)
\end{align}
Photon states span a Hilbert space denoted ${\cal H}_{\rm ph}$, and can be used to write down an arbitrary density operator in a space ${\cal D}_{\rm ph}$. 

The Hilbert space, operator algebra, and density operator space of the composite theory are ${\cal H}={\cal H}_m \otimes {\cal H}_{\rm ph}$, ${\cal A}={\cal A}_m\otimes {\cal A}_{\rm ph}$, and ${\cal D}={\cal D}_m \otimes {\cal D}_{\rm ph}$ respectively. To understand the meaning of the light and matter subsystems it is necessary to determine the meaning of the canonical operators by writing them in terms of operators whose physical meaning is known. 

To this end we begin by noting that, as a postulate of electrodynamics (classical or quantum), the fields ${\bf E}$ and ${\bf B}$ exhaustively assign electromagnetic properties to each individual event in spacetime. These fields are therefore said to be {\em local}. Similarly, ${\bf \rho}$ and ${\bf J}$ assign local material properties via the positions and velocities of charges, such as ${\bf r}$ and ${\dot {\bf r}}$. A function of local fields and of their spacetime derivatives at a single point in space and time is also local. An electromagnetic example is the energy density $[{\bf E}^2+{\bf B}^2]/2$, and a material example is the source of electric waves, ${\bm \mu}=-{\dot {\bf J}}-\nabla\rho$, which is such that $\square {\bf E}(t,{\bf x})={\bm \mu}(t,{\bf x})$. Any functional of local fields that depends on local fields over a finite region of space at the same time, is non-local. Examples include the gauge-invariant transverse potential
\begin{align}
{\bf A}_{\rm T}(t,{\bf x}) = [(\nabla \times )^{-1}{\bf B}](t,{\bf x}) = \int d^3 x' {\nabla'\times {\bf B}(t,{\bf x}') \over 4\pi|{\bf x}-{\bf x}'|},\label{Ati}
\end{align}
and the Coulomb potential and longitudinal electric field
\begin{align}
&\phi(t,{\bf x}) = [-\nabla^{-2}\rho](t,{\bf x}) =  \int d^3 x' {\rho(t,{\bf x}') \over 4\pi|{\bf x}-{\bf x}'|},\\
&{\bf E}_{\rm L}(t,{\bf x})=-\nabla \phi(t,{\bf x}).\label{eli}
\end{align}
A fundamental non-local observable of an atom-field system is the total energy
\begin{align}\label{tote}
H =& {1\over 2}m{\dot {\bf r}}^2+{1\over 2}\int d^3 x \left[{\bf E}({\bf x})^2+{\bf B}({\bf x})^2\right]\nonumber \\
=&{1\over 2}m{\dot {\bf r}}^2+V({\bf r}) + {1\over 2}\int d^3 x \left[{\bf E}_{\rm T}({\bf x})^2+{\bf B}({\bf x})^2\right],
\end{align}
where in writing the second equality we have ignored the infinite Coulomb self-energies of the charges. When written in terms of canonical operators in the Coulomb gauge the energy becomes the Coulomb-gauge Hamiltonian
\begin{align}\label{h}
H =& {1\over 2m}\left[{\bf p}-q{\bf A}_{\rm T}({\bf r})\right]^2+V({\bf r})+  {1\over 2}\int d^3 x \left[{\bf \Pi}({\bf x})^2+{\bf B}({\bf x})^2\right]\nonumber \\
=& H_m+H_{\rm ph}-{q\over m}{\bf p}\cdot {\bf A}_{\rm T}({\bf r})+{q^2\over 2m}{\bf A}_{\rm T}({\bf r})^2,
\end{align}
which it is easily verified yields the correct Maxwell-Lorentz equations. The Hamiltonian can be used to determine the meaning of the canonical momenta ${\bf p}$ and ${\bf \Pi}$ in the Coulomb gauge as
\begin{align}
&m{\dot {\bf r}} = -i[{\bf r},H] = {\bf p}-q{\bf A}_{\rm T}({\bf r}), \\ 
&{\dot {\bf A}}_{\rm T}({\bf x})\equiv -{\bf E}_{\rm T}({\bf x})=-i[{\bf A}_{\rm T}({\bf x}),H]={\bf \Pi}({\bf x}).
\end{align}
 For notational economy the time argument $t$ of Heisenberg picture operators is omitted. Note that when substituted back into Eq.~(\ref{h}) these equations give Eq.~(\ref{tote}) as required. Immediately we see that the photonic momentum is the non-local transverse electric field 
\begin{align}\label{piet}
-{\bf \Pi}({\bf x})&={\bf E}_{\rm T}({\bf x}) = \int d^3 x' \delta^{\rm T}({\bf x}-{\bf x}')\cdot {\bf E}({\bf x}') \nonumber \\ &= {\bf E}({\bf x})  +\nabla \int d^3 x' {\rho({\bf x}')\over 4\pi|{\bf x}-{\bf x}'|}.
\end{align}
The material momentum ${\bf p}$ can be written using Eqs.~(\ref{Ati}) and (\ref{eli}) as
\begin{align}\label{pP}
{\bf p}=m{\dot {\bf r}} + q{\bf A}_{\rm T}({\bf r}) = m{\dot {\bf r}} + P_{\rm long}
\end{align}
where
\begin{align}
P_{\rm long} = \int d^3 x {\bf E}_{\rm L{\bf r}}({\bf x})\times {\bf B}({\bf x})
\end{align}
is the electromagnetic momentum associated with the longitudinal electric field generated by the dynamical charge at ${\bf r}$ with density $\rho_{\bf r}({\bf x})=q\delta({\bf x}-{\bf r})$;
\begin{align}
{\bf E}_{\rm L{\bf r}}({\bf x}) = -\nabla \int d^3 x' {\rho_{\bf r}({\bf x}')\over 4\pi|{\bf x}-{\bf x}'|}\label{elr}.
\end{align}

We have specified precisely what is meant by a {\em quantum system}, and also precisely what is meant by {\em locality}. Notice that the current ${\bf J}$ and associated mechanical momentum $m{\dot {\bf r}}$ are local, but by definition they are not ``material" system observables, that is, they do not possess the form $O\otimes I_{\rm ph}$. The canonical momentum ${\bf p}$ is by definition ``material" but according to Eq.~(\ref{pP}), it may be highly non-local. Similarly the total electric field is local, but according to Eq.~(\ref{piet}) it is certainly not ``photonic". The canonical momentum ${\bf \Pi}$ is by definition photonic, of the form $I_m\otimes O$, but again according to Eq.~(\ref{piet}) it may be highly non-local. It is remarkable that the quantum information-theoretic notion of locality based on tensor-product structure, according to which any unitary operation performed on a subsystem is said to be ``local", is generally different to that based on the field concept. When this is relevant, the terminology ``local operation" used in quantum information theory should be replaced with the term  ``subsystem operation" and it must be recognised that such operations may be highly non-local in space.

In the Coulomb gauge, the matter subsystem is defined not as comprising bare mechanical charges, but as non-local charges dressed by their electrostatic fields. This dressed matter produces transverse electromagnetic waves that propagate at the speed of light via solutions of the wave equation
\begin{align}
&\square {\bf A}_{\rm T}(t,{\bf x}) = {\bf J}_{\rm T}(t,{\bf x}),\label{cwave}
\end{align}
together with ${\bf E}_{\rm T}=-{\dot {\bf A}}_{\rm T}$ and ${\bf B}=\nabla\times {\bf A}_{\rm T}$. Since the wave operator $\square$ can be applied to any suitably differentiable electromagnetic field, it is clear that what distinguishes electromagnetic waves are the sources that produce them, and in Eq.~(\ref{cwave}) ${\bf J}_{\rm T}$ is not a field of a localised source. Indeed, ${\bf J}_{\rm L}=-{\dot {\bf E}}_{\rm L}$ and so ${\bf J}_{\rm T}= {\bf J}+{\dot {\bf E}}_{\rm L}$.
The {\em local} current ${\bf J}$ is that of bare charges, while ${\dot {\bf E}}_{\rm L}$ is the non-local ``current" of the electrostatic dressing field. The field ${\bf E}_{\rm L}$ decays polynomially away from a charge and so strictly speaking, even in a theory that considers point charges, Coulomb gauge ``matter" extends to spatial infinity.

\subsubsection{Multipolar gauge}

A second commonly used gauge in atomic physics is the multipolar gauge, also variously called the Poincar\'e gauge, and the Power-Zienau-Woolley gauge. It defines an atom as a {\em localised} bound charge system acting as a whole, and concurrently defines photons via a causal field produced by these localised sources. 

The general decomposition ${\bf E}=-{\bf \Pi}-{\bf P}$ is obtained by means of a unitary gauge-fixing transformation of the Coulomb gauge theory using
\begin{align}\label{R}
R=\exp\left[-i\int d^3 x {\bf P}({\bf x})\cdot {\bf A}_{\rm T}({\bf x})\right],
\end{align} 
which is such that $R{\bf \Pi} R^\dagger = {\bf \Pi}+{\bf P}_{\rm T}$. The electric field observable is represented in the new frame by ${\bf E} = -R{\bf \Pi}R^\dagger - {\bf P}_{\rm L} =-{\bf \Pi}-{\bf P}$. Since the transformation is not of product form the light and matter subsystems corresponding to each different electric field decomposition are physically distinct, but in every frame purely transverse photons can be defined using ${\bf \Pi}$ via Eq.~(\ref{phot}).

If, unlike in the Coulomb gauge wherein ${\bf P}={\bf P}_{\rm L}$, we define the material part ${\bf P}$ to be localised inside the bare atom, then at all points outside the atom we will have ${\bf P}_{\rm T}=-{\bf P}_{\rm L}={\bf E}_{\rm L}$.  Although we cannot represent ${\bf E}$ exactly using a transverse photonic momentum ${\bf \Pi}$, we will have successfully found a new frame (gauge) in which the physical meaning of the operator ${\bf \Pi}$ is $-{\bf E}_{\rm T}-{\bf P}_{\rm T}=-{\bf E}-{\bf P}$, which equals $-{\bf E}$ at every point outside the atom.

A local choice of ${\bf P}$ becomes immediately apparent by noting that ${\bf P}_{\rm L}=-{\bf E}_{\rm L}$ can be written
\begin{align}\label{long}
{\bf P}_{\rm L}({\bf x}) = q\int_{\bf 0}^{\bf r} d{\bf s} \cdot \delta^{\rm L}({\bf x}-{\bf s}) = \int_0^1 d\lambda {\bf d}\cdot \delta^{\rm L}({\bf x}-\lambda {\bf r})
\end{align}
where ${\bf d}=q{\bf r}$ and ${\bf s} =\lambda{\bf r}, ~0\leq\lambda\leq 1$ is the straight line from the charge $-q$ at the origin to the dynamical charge $q$ at ${\bf r}$. ``Completing" the longitudinal $\delta$-function in Eq.~(\ref{long}) by adding the same line-integral of the transverse $\delta$-function, yields a much more localised solution of Gauss' law first used by Power and Zienau \cite{power_coulomb_1959}, namely 
\begin{align}
{\bf P}({\bf x}) = q\int_{\bf 0}^{\bf r} d{\bf s}\delta({\bf x}-{\bf s}) = \int_0^1 d\lambda {\bf d}\delta({\bf x}-\lambda {\bf r}).
\end{align}
The Power-Zienau-Woolley transformation is defined by using this choice of ${\bf P}$ in Eq.~(\ref{R}). The meaning of the light and matter subsystems in the multipolar gauge can be deduced using the multipolar Hamiltonian
\begin{align}
H'=RHR^\dagger =& {1\over 2m}\left[{\bf p}-q{\bf A}({\bf r})\right]^2+V({\bf r})\nonumber \\ &+  {1\over 2}\int d^3 x \left[[{\bf \Pi}({\bf x})+{\bf P}_{\rm T}({\bf x})]^2+{\bf B}({\bf x})^2\right]
\end{align}
where
\begin{align}\label{pota}
{\bf A}({\bf x}) = -\int_0^1 d\lambda \lambda{\bf x}\times {\bf B}(\lambda{\bf x})
\end{align}
is the Poincar\'e gauge potential satisfying ${\bf x}\cdot {\bf A}({\bf x})=0$. The Heisenberg equation yields the expected result ${\dot {\bf A}}_{\rm T} = -{\bf E}_{\rm T} = {\bf \Pi}+{\bf P}_{\rm T}$ such that ${\bf \Pi}$ now represents a much more localised observable, ${\bf \Pi}=-{\bf E}_{\rm T}-{\bf P}_{\rm T}=-{\bf E}-{\bf P}$. The mechanical momentum is found similarly to be $m{\dot {\bf r}}={\bf p}-q{\bf A}({\bf r})$ so that ${\bf p}=m{\dot {\bf r}}+q{\bf A}({\bf r})$. Eq.~(\ref{pota}) reveals that ${\bf A}({\bf r})$ depends on the local magnetic field only at points inside the atom, implying that ${\bf p}$ also represents a much more localised observable in this gauge.

The multipolar gauge is amenable to a multipole expansion that mirrors classical multipolar radiation theory. To the leading (electric dipole) order, the multipolar polarisation field is that of a dipole at the origin; ${\bf P}({\bf x})={\bf d}\delta({\bf x})$. At this order, the potential ${\bf A}$ vanishes at the point ${\bf r}$, so the material canonical momentum is bare, ${\bf p}=m{\dot {\bf r}}$. We have arrived at a theory in which ``matter" is a fully localised bare dipole, and ``photons" are defined using a field ${\bf \Pi}$ which equals the local electric field $-{\bf E}$ except at the dipole's position ${\bf 0}$. We note that the Coulomb gauge theory in the electric dipole approximation (EDA) is obtained by replacing ${\bf A}_{\rm T}({\bf r})$ with ${\bf A}_{\rm T}({\bf 0})$ in Eq.~(\ref{h}). It is connected to the dipole gauge through 
\begin{align}\label{edaR}
R=e^{-i{\bf d}\cdot {\bf A}_{\rm T}({\bf 0})},
\end{align}
which is the EDA of $R$ in Eq.~(\ref{R}).

\subsection{Gauge-nonrelativistic predictions}\label{elim}

We have seen that light and matter subsystems in QED can only be assigned physical meaning relative to a choice of gauge. Unitary gauge-fixing transformations are not of product form. It is however, important to consider whether and when this relativity can be ignored. In special relativity, for example, when the relative velocity between inertial frames $A$ and $B$ is sufficiently small, the associated space and time intervals between any two events are approximately the same; $\Delta x_A \approx \Delta x_B$ and $\Delta t_A \approx \Delta t_B$. The relativity of space and time, i.e., the mixing effect of the Lorentz transformation $\Lambda$ from $A$ to $B$, can then be ignored completely, such that space and time emerge as approximately absolute (frame-independent) concepts, as is assumed from the outset in Galilean relativity. 

Consider the case of spontaneous emission of a dipole. The eigenvector $\ket{e,0}$ of $h=H_m+H_{\rm ph}$ with an excited dipole and no photons represents a different physical state in the Coulomb and multipolar gauges. If we denote the state that it represents in the Coulomb (multipolar) gauge by ${\cal S}$ (${\cal S}'$), then in the multipolar (Coulomb) gauge this same state is represented by $R\ket{e,0}$ ($R^\dagger \ket{e,0}$), which is entangled. Physically, ${\cal S}'$ is an excited state of a bare dipole, whereas ${\cal S}$ is an excited state of an electrostatically dressed dipole. Similarly, the vector $\ket{g,{\bf k}\lambda}$ with a ground level dipole and a photon of momentum ${\bf k}\lambda$ represents different physical states ${\cal F}$ and ${\cal F}'$ in the Coulomb and multipolar gauges respectively. 


However, if we suppose that the atom-field interaction vanishes in the remote past $t=-\infty$ and the distant future $t=+\infty$, then at these times $h=H$ and each unperturbed eigenvector uniquely represents a physical state (the transformation $R$ becomes the identity). If the interaction is switched on and then off over an infinite amount of time (a scattering process) then we can define the total rate of spontaneous emission for the transition $e\to g$ by
\begin{align}
\Gamma_{eg} =2\pi \int d^3 k \sum_\lambda |\bra{g,{\bf k}\lambda}V\ket{e,0}|^2\delta(\omega-\omega_{eg}) = {\omega_{eg}^3|{\bf d}_{eg}|^2\over 3\pi}.
\end{align}
Here $\omega=|{\bf k}|$ is the energy of the emitted photon, $\omega_{eg}=\omega_e-\omega_g$ is the energy of the atomic transition, and the $\delta$-function imposes the condition that these energies are the same. The rate $\Gamma_{eg}$ is the same whether the interaction $V=H-h$ of the Coulomb gauge is used, or the interaction $V'=H'-h$ of the multipolar gauge is used. 

This gauge non-relativistic property of the QED $S$-matrix is general \cite{stokes_implications_2022,cohen-tannoudji_photons_1989,woolley_gauge_2000,craig_molecular_1998}, but it must be distinguished from gauge-invariance, which is merely a special case of Eq.~(\ref{gi}). The amplitude of the process ${\cal S}\to {\cal F}$ is immediately calculable in the Coulomb gauge as $b_{{\bf k}\lambda;eg} = \bra{g,{\bf k}\lambda}U(t)\ket{e,0}$ and the same prediction is obtained in the multipolar gauge as $b_{{\bf k}\lambda;eg} =\bra{g',({{\bf k}\lambda})'}U'(t)\ket{e',0'}$ where $\ket{\psi'}=R\ket{\psi}$ and $U'=RUR^\dagger$. This prediction is therefore gauge-invariant. Similarly, the amplitude $b'_{{\bf k}\lambda;eg}=\bra{g,{{\bf k}\lambda}}U'(t)\ket{e,0}$ of the process ${\cal S}'\to {\cal F}'$ is a gauge-invariant prediction. Under the conditions that define the $S$-matrix, the total rates associated with these generally distinct physical processes are actually the same. In this context we can speak of ``spontaneous emission" as a unique process. 

Another important example of a gauge non-relative prediction is the quantum optical Lindblad master equation \cite{stokes_implications_2022}.  
In general the master equations ${\dot \rho}_m(t) = i{\rm tr}_{\rm ph}[\rho(t),H]$ and ${\dot \rho}'_m(t) = i{\rm tr}_{\rm ph}[\rho'(t),H']$ describe the dynamics of physically distinct atomic subsystems. However, letting $H_m=\sum_n \omega_n \ket{n}\bra{n}$ and assuming for simplicity that the $\omega_n$ are non-degenerate, the Born-Markov and secular approximations mimic the assumptions that define the $S$-matrix, such that both master equations reduce to the same Lindblad equation of the form
\begin{align}
{\dot \rho}_m(t) =- i[H_m+H_{\rm LS},\rho(t)] + {\cal D}[\rho_m(t)]\label{me}
\end{align}
where
\begin{align}
{\cal D}(\rho_m) := \sum_{n,p,q,r \atop \omega_{np} =\omega_{qr}>0} \gamma_{npqr}\left(L_{np}\rho L^\dagger_{qr} -{1\over 2}\left\{L^\dagger_{qr}L_{np},\rho\right\} \right)\label{diss}
\end{align}
with $L_{np}:=\ket{p}\bra{n}$, and
\begin{align}
H_m+H_{\rm LS} = \sum_n ({\epsilon_n+\Delta_n})\ket{n}\bra{n}.\label{LS}
\end{align}
The rates $\gamma_{npqr}$ and shifts $\Delta_n$ are, like $\Gamma_{eg}$, defined in terms of the unperturbed atom-photon states, but they are likewise independent of the chosen interaction Hamiltonian, because they are on-energy-shell matrix elements. The approximations that lead to the final result in Eq.~(\ref{me}) can therefore be said to define the gauge non-relativistic regime \cite{stokes_implications_2022}.

\section{Gauge relativity of information in toy models}\label{toy}

\subsection{Motivation}\label{mot}

The significance of subsystem (gauge) relativity in QED is brought into focus when considering the most fundamental aspects of energy and information exchange between multiple material systems. Current interest in such topics extends beyond the confines of QED itself. A consideration of electromagnetic interactions has been used to shed light on the quantum nature of gravitational interactions. A question of major importance is whether confirmation of entanglement between two masses can be taken to infer the quantisation of gravity and the existence of mediating quanta (gravitons) \cite{marletto_gravitationally_2017,bose_spin_2017,fragkos_inference_2022}. 

The simplest scenario consists of two stationary hydrogen atoms at ${\bf R}_1={\bf 0}$ and ${\bf R}_2={\bf R}$ with charge density 
\begin{align}
\rho({\bf x}) =& \rho_1({\bf x})+\rho_2({\bf x}) \nonumber \\ \equiv & q[\delta({\bf x}-{\bf r}_1)-\delta({\bf x})]+q[\delta({\bf x}-{\bf r}_2)-\delta({\bf x}-{\bf R})].
\end{align}
The Coulomb gauge Hamiltonian is
\begin{align}
H=H_m^1+H_m^2+H_{\rm ph}+V^1+V^2+ V^{\rm inter}
\end{align}
where $H_m^i$ and $V^i$ are the individual atom Hamiltonians and atom-field interactions respectively, defined for each atom as in Eq.~(\ref{h}). The additional term now appearing
\begin{align}\label{coul2}
V^{\rm inter} = \int d^3 x\, {\bf P}_{\rm L1}({\bf x}) \cdot {\bf P}_{\rm L2}({\bf x}) 
\end{align}
where ${\bf P}_{{\rm L}i}=-\nabla(\nabla^{-2}\rho_i)$, is the interatomic Coulomb interaction. It is the energy contained in the overlapping electrostatic dressing fields of two non-local atoms defined relative to the Coulomb gauge. It results in correlations between the atoms even before a causal signal can traverse the distance $R$, and is of course the direct electromagnetic analog of the Newtonian gravitational interaction between masses.

We see that photon mediated interactions are not the only possibly cause of entanglement between atoms defined relative to the Coulomb gauge, because such atoms are extended objects that overlap. It should be noted however, that a Coulomb gauge theory without photons, that is, one that includes only the term $V^{\rm inter}$, is incomplete and exhibits important kinematical differences. In particular, the atomic canonical momenta within such a theory are the localised mechanical momenta, ${\bf p}_i=m{\dot {\bf r}}_i$, whereas within the complete theory these momenta are given by Eq.~(\ref{pP}).

In the multipolar gauge the material part of the electric field is a sum of multipolar polarisation fields for each atom, ${\bf P}={\bf P}_1+{\bf P}_2$, and each component is localised inside the corresponding atom. Unlike Eq.~(\ref{coul2}), the overlap integral between ${\bf P}_{1}$ and ${\bf P}_{2}$ vanishes, so there is no direct interatomic interaction in the multipolar Hamiltonian, which is of the form
\begin{align}
H'=RHR^\dagger =H_m^1+H_m^2+ H_{\rm ph}+V'^1+V'^2.
\end{align}
All interactions are mediated by a causal field that equals the local electric field outside the localised atoms defined relative to this gauge. Correlations between these atoms cannot occur in the absence of photons defined relative to the multipolar gauge.

The multipolar and Coulomb gauges are of course equivalent and all predictions obtained from the theory are gauge-invariant. Beyond scattering theory, however, questions concerning the physics of atoms and photons are ill-posed without an unambiguous specification of the physical degrees of freedom that have been used to define these concepts. Two opposite (but not contradictory) conclusions regarding the necessary conditions for entanglement to arise between atoms have been arrived at above because the atoms and therefore the entanglement referred to within these conclusions is different.

It must be recognised that entanglement occurs between {\em observables}. An eigenvector of an operator representing an observable represents a state in which the observable possesses the corresponding eigenvalue with certainty. Eigenvectors provide orthonormal bases for the Hilbert state space of a quantum system. When expanded in such a basis the vector representing a given quantum state may or may not be entangled. The term ``atom" or ``photon" should merely be considered an occasionally convenient abbreviation for a certain observable or collection of observables. The latter can always be specified in terms of primitive local and manifestly gauge-invariant fields whose physical meaning is immediate, and which are not defined in terms of prior concepts (save for spacetime itself). Examples are given by Eqs.~(\ref{piet})-(\ref{elr}).

A mediating field, by which we mean the field of a non-material (photonic) quantum subsystem, is necessary to produce entanglement between localised atomic quantum subsystems. Yet, such a mediating field is unnecessary to produce entanglement between non-local (overlapping) atomic quantum subsystems. Only an analysis of the specific operational procedures used in an experiment can determine which {\em observables} are accessed and thereby which definitions of atoms and photons are relevant. As we remarked at the outset, the relevant partition of a quantum system into subsystems is determined by the operationally available interactions and measurements. In practice any operational procedure will possess finite extent in space and time, which one might plausibly suggest must somehow be correlated with the spatio-temporal localisation properties of the quantum system it grants access to. 

It should be noted however, that distinct definitions of atom are not necessarily independent. A change in the degree of entanglement between non-local atoms will generally imply a change in the degree of entanglement between localised atoms. Moreover, different definitions of atom should not be considered indistinguishable experimentally. Indeed, it is in terms of the {\em observables} that define them that any two conceptions of atom differ. Like gauge invariance, subsystem gauge relativity is fundamental. Gauge non-relative predictions of atomic and photonic properties, that is, those independent of the considered observables, will generally only be obtained when idealisations (as in scattering theory) and approximations (such as Born-Markov and secular/rotating-wave approximations) are employed.  

Even within scattering theory subsystem relativity is apparent. The matrix element describing resonance energy transfer in the vacuum between identical dipoles that are respectively in excited and ground states represented by $\ket{e_1}$ and $\ket{g_2}$, is to second order in the interaction given by
\begin{align}\label{M}
M=\bra{f}V\ket{i} + \sum_I {\bra{f}V\ket{I}\bra{I}V\ket{f} \over E_i-E_I}
\end{align}
where $\ket{f}=\ket{g_1,e_2,0}$, $\ket{i}=\ket{e_1,g_2,0}$, and $E_n$ is the energy associated with the state represented by $\ket{n}$. The sum in Eq.~(\ref{M}) extends over all intermediate eigenvectors of $h=H_m^1+H_m^2+H_{\rm ph}$. In the dipole gauge, the first term in Eq.~(\ref{M}) does not contribute, because there are no direct interatomic interactions between localised dipoles. From the second term in Eq.~(\ref{M}) one obtains (latin indices denote cartesian components and repeated indices are summed)
\begin{align}\label{M'}
M = {\omega_{eg}^3 d_id_j \over 4\pi}\left[\beta_{ij}\left({\cos\xi \over \xi^3}+{\sin\xi \over \xi^2}\right)-\gamma_{ij} {\cos \xi \over \xi}\right]
\end{align}
where $\xi=\omega_{eg}R$, $\beta_{ij}=\delta_{ij}-3{\hat R}_i{\hat R}_j$, and $\gamma_{ij}=\delta_{ij}-{\hat R}_i{\hat R}_j$. In the Coulomb gauge the first term in Eq.~(\ref{M}) is non-zero due to the overlap $V^{\rm inter}$, which however, does not contribute to the second term in Eq.~(\ref{M}). The latter is found to be $M-\bra{f}V^{\rm inter}\ket{i}$ where $M$ is given in Eq.~(\ref{M'}) \cite{craig_molecular_1998}. In the Coulomb gauge then, the atom-field interaction implicitly includes a static contribution that exactly cancels the contribution of the inter-dipole electrostatic overlap.

Energy transfer involving the same initial and final states was considered early on by Fermi, in order to demonstrate that inter-atomic interactions via photon exchange do not violate causality \cite{fermi_quantum_1932}. Such dynamical considerations are beyond the scope of the stationary perturbation theory giving Eq.~(\ref{M}). Fermi considered atoms that were not directly coupled and obtained a causal result for the excitation probability at time $t$ of the initially unexcited atom. The problem received renewed interest much later on however, when it was noted that Fermi made Markovian type approximations without which an {\em apparently} acausal result would be obtained \cite{cohen-tannoudji_photons_1989,biswas_virtual_1990,milonni_photodetection_1995,power_analysis_1997,buchholz_there_1994,hegerfeldt_causality_1994}. Indeed, an (in)famous result due to Hegerfeldt \cite{hegerfeldt_causality_1994} found that under very mild assumptions, the probability of excitation of the second atom is necessarily non-zero for times less than $R$. The apparent contradiction is resolved by noting that even a single atom in the vacuum immediately possesses a non-zero excitation probability due to virtual photon absorption; the vector $\ket{g,0}$ is not an eigenvector of the total atom-field Hamiltonian. In the two atom problem the excitation probability of atom 2 at time $t$ can be partitioned as
\begin{align}\label{prob}
P(t) = P_0(t)+P_1(t)
\end{align}
where $P_0$ and $P_1$ are independent and dependent on atom 1 respectively. If the dipoles envisaged by Fermi are defined relative to the multipolar gauge then $P_1(t)$ is necessarily zero for $t<R$ as required by causality. This becomes immediately apparent by noting that the field ${\bf \Pi}$ to which the dipole 2 couples at it's own position ${\bf R}$, is a superposition ${\bf \Pi}={\bf \Pi}_2+{\bf \Pi}_{\rm vac}+{\bf \Pi}_1$ consisting of it's own reaction field, the vacuum field, and the source field ${\bf \Pi}_1$ of atom $1$. The contribution $P_1(t)$ in Eq.~(\ref{prob}) results directly from the source field ${\bf \Pi}_1$. In the multipolar gauge this source field at any point ${\bf x}\neq {\bf 0}$ (including ${\bf x}={\bf R}$) is (minus) the well-known causal electric source field of a point dipole \cite{power_time_1999};
\begin{align}
-{\bf \Pi}_1(t,{\bf x})={\bf E}_s(t,{\bf x}) = \theta(t_r)\nabla\times \left[\nabla \times {{\bf d}(t_r) \over 4\pi x}\right]\label{elec0}
\end{align}
where $t_r=t-x$, $\theta$ is the Heaviside step function, and ${\bf d}$ is the dipole moment operator of dipole $1$. Clearly dipole 1 cannot influence dipole 2 for times $t<R$. This same conclusion holds at every multipole order, i.e., for atoms that are not approximated as dipoles.

If one considers dipoles defined relative to the Coulomb gauge then dipole 2 immediately exhibits a non-zero excitation probability that does depend on dipole 1, behaviour that lies in stark contrast to that of dipoles defined relative to the dipole gauge. The kind of exact cancellation of terms that occurs in the Coulomb gauge to give Eq.~(\ref{M'}) and thereby eliminate gauge-relativity, can only generally occur within the confines of stationary perturbation theory. The immediate excitation occcuring in the Coulomb gauge is however not a violation of causality, because the atoms involved cannot  strictly be considered spatially disjoint. The term $V^{\rm inter}$ appearing in Eq.~(\ref{coul2}) is what describes their overlap. Indeed, the electric source field produced by a single atom defined relative to the Coulomb gauge is
\begin{align}
{\bf E}_s(t,{\bf x}) =\theta(t_r)\nabla\times \left[\nabla \times {{\bf d}(t_r) \over 4\pi x}\right] + \theta(-t_r){\bf E}_{\rm L}(0,{\bf x}),\label{e2}
\end{align}
which is not the same as Eq.~(\ref{elec0}). This difference is not contradictory, because the total electric field is ${\bf E}(t,{\bf x}) = {\bf E}_{\rm vac}(t,{\bf x})+{\bf E}_s(t,{\bf x})$ and the vacuum parts ${\bf E}_{\rm vac}(t,{\bf x})$ differ between the two gauges by the exact negative of the difference in source parts. The total field ${\bf E}$ is unique (gauge non-relative) but its partitioning into vacuum (photon) and source (atom) parts is, as expected, gauge-relative. The final term $\theta(-t_r){\bf E}_{\rm L}(0,{\bf x})$ in Eq.~(\ref{e2}) again demonstrates that Coulomb gauge matter includes the electrostatic field by definition.

One might conclude that atoms defined relative to the Coulomb gauge are somewhat unphysical, but it should be born in mind that for sufficiently large $R$ the overlap $V^{\rm inter}$ will become exceedingly small and negligible in practice. Operationally, it is also not clear that the opposite extreme of point-localised dipoles, as are defined relative to the multipolar gauge, are necessarily the most relevant definition to consider. For these dipoles, returning to Eq.~(\ref{prob}), Hegerfeldt's theorem states that $P(t)\neq 0$ for $t<R$, which implies that a nonzero $P_0(t)$ is necessary to preserve causality between localised dipoles, and yet such immediate vacuum excitations may also be viewed as unphysical. A successful renormalisation procedure might be viewed as one that absorbs these virtual excitations within a new definition of the subsystems for which $\ket{g,0}$ is the true ground state of $H$, in which case $P_0(t)=0$. For simple dipolar systems, such a gauge does indeed exist and possesses an interaction Hamiltonian that symmetrically mixes the linear Coulomb and multipolar gauge couplings \cite{stokes_implications_2022,drummond_unifying_1987,baxter_gauge_1990,stokes_gauge_2019,stokes_extending_2012,stokes_alternative_2015,stokes_gauge_2013}. The resulting definition of matter must be non-local otherwise $P_0(t)=0$ would imply a violation of causality. Indeed, this definition of matter incurs a level of material dressing that is in between that of the Coulomb gauge (total electrostatic dressing) and the dipole gauge (no dressing). This is discussed further in Sec.~\ref{randd}

It is clear that subsystem gauge relativity is a fundamental feature that cannot in general be ignored. The observables defining light and matter quantum subsystems are in general non-local, as is revealed by expressing them in terms of manifestly {\em gauge-invariant} and {\em local} fields. It is not immediately obvious which definitions of quantum subsystems are the most operationally relevant, and how the answer to this question might depend on experimental context. Motivated by this observation, below we turn our attention toward understanding the different behaviours of energy and information when different subsystem definitions are adopted. To this end, we focus on two simple and transparent toy models.

\subsection{One dipole and one mode}

\subsubsection{Model}\label{JC}

If we suppose that a dipole-field system is enclosed inside a box with prescribed boundary conditions then the allowed photonic wavevectors become discrete. In the multipolar gauge, because the photonic field is locally defined, the boundaries can be described entirely using appropriate mode-functions $\{{\bf f}_{\bf k\lambda}({\bf x})\}$, without having to introduce image charges \cite{power_quantum_1982,vukics_adequacy_2012}. We therefore begin in this gauge and assume that the dipole couples appreciably to only one photonic mode with polarisation vector ${\bf e}$ and wavevector ${\bf k}$, tuned such that $\omega =|{\bf k}|$ is very near to the lowest atomic transition energy. Assuming that the dipole is located at a field maximum, ${\bf f}({\bf 0})={\bf e}$, we can write the photonic canonical momentum at this position as \cite{hinds_cavity_1990}
\begin{align}\label{canmom}
{\bf \Pi}({\bf 0}) = i\sqrt{\omega \over 2v}{\bf e}a^\dagger + {\rm H.c.}
\end{align}
where $[a,a^\dagger]=1$, and
\begin{align}
v = \int_{\rm box} d^3 x |f({\bf x})|^2
\end{align}
is the effective mode volume with $f({\bf x})={\bf e}\cdot {\bf f}({\bf x})$. In a periodic cavity, for example, we have $f({\bf x})=e^{i{\bf k}\cdot {\bf x}}$.

In practice multiple modes may need to be considered in order to obtain an accurate model. In principle, additional modes could be included one-by-one until convergence is reached within the envelope of a natural cut-off function that is consistent with the EDA and the non-relativistic treatment of charges. The single mode theory has the benefit of simplicity while possessing essentially the same physical structure, so it is better suited to our purpose. If we assume for further simplicity that ${\bf d}\cdot {\bf e} = d$, then we obtain an essentially one-dimensional model
\begin{align}
H'=H_m+H_{\rm ph}+V'
\end{align}
where
\begin{align}
H_m=& {{\bf p}^2\over 2m}+V({\bf r}) = \sum_n \omega_n \ket{n}\bra{n},\label{smbhm}\\
H_{\rm ph} =& \omega \left(a^\dagger a+{1\over 2}\right),\label{smph}\\
V' =& {d^2  \over 2v} + i d\sqrt{\omega\over 2v}(a^\dagger -a) \equiv {\epsilon}^{\rm dip}+{\tilde V}'.\label{smint}
\end{align}
A corresponding model in an arbitrary gauge controlled using a parameter $\alpha$  can be constructed as $H_\alpha=R_{1\alpha} H'R_{1\alpha}^\dagger$ where [cf. Eq.~(\ref{edaR})]
\begin{align}
R_{\alpha \alpha'}= e^{i(\alpha-\alpha') {\bf d}\cdot {\bf A}_{\rm T}({\bf 0})}
\end{align}
is a gauge-fixing transformation from the gauge $\alpha$ to the gauge $\alpha'$, in which
\begin{align}
{\bf A}_{\rm T}({\bf 0}) = {{\bf e}a^\dagger\over \sqrt{2\omega v}} + {\rm H.c.}
\end{align}
is conjugate to ${\bf \Pi}({\bf 0})$ in Eq.~(\ref{canmom}). The extremal choices $\alpha=0$ and $\alpha=1$ yield the Coulomb and dipole gauges respectively.

If $\rho'=\rho_1$ represents the state of the system in the dipole gauge then $\rho_\alpha=R_{1\alpha} \rho_1 R_{1\alpha}^\dagger$ represents the same state in the gauge $\alpha$. The reduced states of the atomic and photonic subsystems defined relative to the gauge $\alpha$ are then represented by $\rho_{\alpha m} = {\rm tr}_{\rm ph}\rho_\alpha$ and $\rho_{\alpha \rm ph}={\rm tr}_m \rho_\alpha$ respectively. The Von Neumann entropy of a density matrix $\rho$ is defined by $S(\rho)=-\rho \ln \rho$. The subsystem entropy $S_{E\alpha} = S(\rho_{\alpha m})=S(\rho_{\alpha \rm ph})$ quantifies the degree of light-matter entanglement in any pure state of the composite. 
We consider the ground state represented in the gauge $\alpha$ by a vector $\ket{G_\alpha}$.

\subsubsection{Results and discussion}\label{randd}

To make contact with the underlying physics discussed in Sec.~\ref{mot} it is useful to isolate the contribution of the static field to the quantities of interest. Let us begin by considering the evolution of the photon annihilation operator $a_{{\bf k}\lambda}$ in a periodic cavity with mode functions ${\bf f}_{\bf k \lambda}({\bf x}) = {\bf e}_{\bf k \lambda}e^{i{\bf k}\cdot {\bf x}}/\sqrt{2v}$. Assuming a dipole source at ${\bf 0}$, in the gauge $\alpha$ integration of the Heisenberg equation ${\dot a}_{{\bf k}\lambda} = -i[a_{{\bf k}\lambda},H_\alpha]$ yields a solution that is the sum of free (vacuum) and source parts as
\begin{align}\label{ca}
&a_{{\bf k}\lambda}(t) = a_{{\bf k}\lambda}(0)e^{-i\omega t}+a^\alpha_{{\bf k}\lambda,s}(t),\\
&a^\alpha_{{\bf k}\lambda,s}(t) = (1-\alpha)a^0_{{\bf k}\lambda,s}(t)+\alpha a^1_{{\bf k}\lambda,s}(t),\label{asalph}
\end{align}
with
\begin{align}
a^0_{{\bf k}\lambda,s}(t)&={i\over \sqrt{\omega}}\int_0^t dt' e^{i\omega(t'-t)} {\bf f}_{{\bf k}\lambda}({\bf 0})^* \cdot {\dot {\bf d}}(t')\label{cs}
\end{align}
and
\begin{align}\label{da}
a^1_{{\bf k}\lambda,s}(t)=\sqrt{\omega}\int_0^t dt' e^{i\omega(t'-t)} {\bf f}_{{\bf k}\lambda}({\bf 0})^* \cdot {\bf d}(t').
\end{align}
Upon integrating Eq.~(\ref{cs}) by parts the Coulomb and dipole gauge source operators are found to be related as 
\begin{align}\label{intbp}
a^0_{{\bf k}\lambda,s}(t)-a^1_{{\bf k}\lambda,s}(t)=&{i\over \sqrt{\omega}} {\bf f}^*_{\bf k\sigma}({\bf 0}) \cdot\left[{\bf d}(t)-{\bf d}(0)e^{-i\omega t}\right] \nonumber \\ =:& \Delta_{{\bf k}\lambda}(t),
\end{align}
from which it follows using Eq.~(\ref{asalph}) that
\begin{align}
&a^\alpha_{{\bf k}\lambda,s}(t) =a^1_{{\bf k}\lambda,s}(t)+(1-\alpha)\Delta_{{\bf k}\lambda}(t).\label{stat}
\end{align}
When substituted into the mode expansion for ${\bf \Pi}$ the multipolar source operator $a^1_{{\bf k}\lambda,s}(t)$ gives, in the mode continuum limit, minus the well-known causal electric source field of a point dipole, Eq.~(\ref{elec0}). Unlike $a^1_{{\bf k}\lambda,s}(t)$, the Coulomb gauge source operator $a^0_{{\bf k}\lambda,s}(t)$ is singular at $\omega=0$ and its singular part is nothing but the difference $\Delta_{{\bf k}\lambda}(t)$, which gives additional static contributions. Substituting $a^0_{{\bf k}\lambda,s}(t)$ into the mode expansion for ${\bf \Pi}$ gives minus the transverse electric source field. 
Eq.~(\ref{e2}) implies that the difference between the Coulomb gauge and multipolar source parts of ${\bf \Pi}$ is therefore $-{\bf E}_{\rm L}(t,{\bf x})+\theta(-t_r){\bf E}_{\rm L}(0,{\bf x})$. These two electrostatic terms result from the ${\bf d}(t)$-dependent and ${\bf d}(0)$-dependent parts of $\Delta_{{\bf k}\lambda}(t)$ respectively. Eq.~(\ref{stat}) shows that in the gauge $\alpha$ they are weighted by $1-\alpha$. By continuously varying $\alpha$ from $1$ to $0$ we vary the degree of electrostatic dressing from no dressing to full dressing.

The difference $\Delta_{{\bf k}\lambda}(t)$ vanishes if we make the perturbative ansatz
\begin{align}
{\bf d}(t) = \sum_{nm} {\bf d}_{nm} \ket{n}\bra{m} e^{i\omega_{nm}t},\label{pert}
\end{align}
and we assume the resonance condition $\omega_{mn}=\omega_k$. This imposes bare energy conservation directly within the integrated equation of motion [Eq.~(\ref{ca})]. More generally, within the perturbative approximation, Eq.~(\ref{pert}), separation of the static contribution as in Eq.~(\ref{intbp}) is equivalently achieved using the identity
\begin{align}
&{1\over \sqrt{\omega}}{\omega_{nm}\over \omega+\omega_{nm}}= {1\over \sqrt{\omega}}-{\sqrt{\omega}\over \omega+\omega_{nm}}\label{id1}
\end{align}
within the Coulomb gauge operator. The second term on the right-hand side gives the multipolar source operator while the remaining term gives the static (singular) part $\Delta_{{\bf k}\lambda}(t)$, which varies as $1/\sqrt{\omega}$. In this way, we can isolate within each mode of the field, the static contributions that serve to delocalise the dipole quantum subsystem. Precisely this kind of separation was first used by Power and Zienau \cite{power_coulomb_1959}. In so doing they discovered that what is now known as multipolar QED (PZW theory) provides a photonic subsystem whose canonical momentum equals the local electric source field outside the atom. 


Returning to the single mode theory, let us now restrict our attention to the weak-coupling regime and also to a single dipolar transition with frequency $\omega_m$ and real transition dipole moment $d_{eg}$. The ground state vector of the dipole-mode system in the gauge $\alpha$ is readily obtained using second order perturbation theory. The reduced material density operator is found subsequently to be diagonal in the bare dipole basis and reads
\begin{align}
\rho_{\alpha m} =& \beta^2_\alpha \ket{e}\bra{e}+(1-\beta^2_\alpha)\ket{g}\bra{g},\label{p}\\
\beta_\alpha=& \eta \left[{\omega_m \over \omega+\omega_m}-\alpha\right] = (1-\alpha)\beta_0+ \alpha \beta_1\label{beta}.
\end{align}
Here $\eta = d_{eg} /\sqrt{2\omega v}$ is a dimensionless coupling parameter, and $p_\alpha=\beta_\alpha^2$ is the population of the excited state. Note that $\eta$ is singular as $1/\sqrt{\omega}$ at $\omega=0$, so $\beta_0$ is similarly singular. We can again use the identity~(\ref{id1}) to separate off the singular part of $\beta_\alpha$ [cf. Eq.~(\ref{stat})] as $\beta_\alpha=\beta_1+s_\alpha$ in which the first term $\beta_1$ gives the non-singular excitation probability $p_1=\beta_1^2$ of a bare localised dipole, and the second term $s_\alpha = (1-\alpha)\eta$ is the purely static deviation from this excitation, which is weighted by $1-\alpha$. Note that, at resonance $\beta_\alpha=\eta(1/2-\alpha)$ and for $\alpha=0$ we obtain $s_0 = -2\beta_1 = \eta$, that is, the static contribution is minus twice the ``local" contribution. As a result $\beta_0=-\beta_1$, so the Coulomb and dipole gauge excited state populations are equal in the perturbative limit; $p_0=p_1$.

A non-zero probability $p_\alpha= \beta^2_\alpha$ in Eq.~(\ref{p}) results from absorption of a photon from the virtual photon cloud dressing the bare dipole in the composite ground state. A gauge which yields $\beta_\alpha \equiv 0$ is given by 
\begin{align}
\alpha = \alpha_{\rm JC}:= {\omega_m\over \omega+\omega_m}={1\over 1+\delta}
\end{align}
where $\delta=\omega/\omega_m$. This choice results in a Jaynes-Cummings model without performing the rotating-wave approximation \cite{stokes_implications_2022,drummond_unifying_1987,baxter_gauge_1990,stokes_gauge_2019,stokes_extending_2012,stokes_alternative_2015,stokes_gauge_2013}. The ground state of this model is the bare ground state $\ket{0,g}$. The two-level dipole defined relative to this gauge can be interpreted, within the approximations made, as subsuming the ground state virtual photon cloud surrounding the bare dipole. At resonance the value of $\alpha_{\rm JC}$ is simply $1/2$. More generally, the $\alpha_{\rm JC}$-gauge (Jaynes-Cummings gauge) symmetrically mixes the Coulomb and dipole gauge couplings. For a mode with wavelength $\lambda \ll 2\pi/\omega_m$, $\alpha_{\rm JC}$ approaches $0$, so with respect to such a mode the virtual cloud appears as a static dressing field. For a mode with $\lambda \gg 2\pi/\omega_m$,  $\alpha_{\rm JC}$ approaches $1$, so the virtual cloud is relatively devoid of photons with such wavelengths, suggesting some degree of localisation.

\begin{figure}[t]
\centering
\hspace*{-2.4mm}
\includegraphics[scale=0.805]{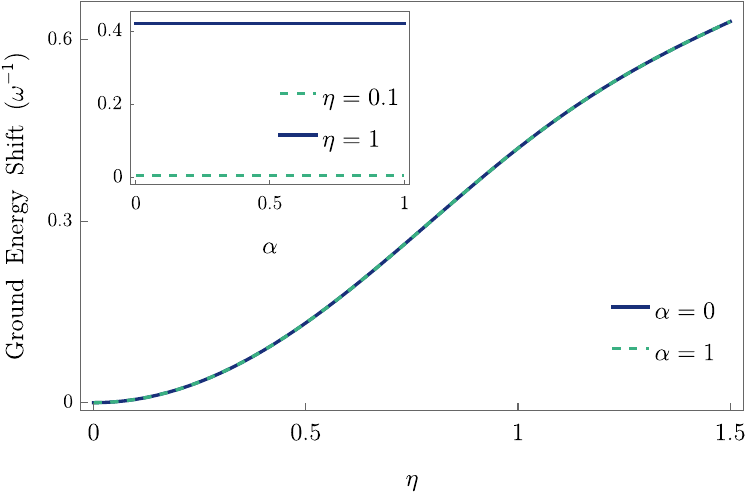}
\vspace*{-0.3cm}
\caption{The ground state energy shift, $E_G(\eta)-E_G(0)$, where $E_G$ is the ground state energy, is plotted as a function of $\eta$ in the Coulomb and multipolar gauges. The inset shows a the same energy as function of $\alpha$ for two values of $\eta$.}
\label{f0}
\vspace*{-0.3cm}
\end{figure}

As an example we consider a double-well dipole with potential $V(\theta,\phi) = -\theta r^2/2 + \phi r^4/4$ where $\theta$ and $\phi$ control the shape of the double-well. The bare Hamiltonian is \cite{de_bernardis_breakdown_2018}
\begin{align}
H_m = {{\cal E}\over 2}\left(-\partial_\zeta^2 -\iota\zeta^2+{\zeta^4 \over 2} \right)
\end{align}
where we have defined dimensionless dipolar operator $\zeta = r/r_0$ with $r_0=(1/[m\phi])^{1/6}$, along with ${\cal E}=1/(mr_0^2)$ and $\iota=\theta m r_0^4$. The difference between the ground energy of $H_\alpha$ and that of $h=H_m+H_{\rm ph}$ is plotted in Fig.~\ref{f0}, which is independent of $\alpha$. Different gauges have therefore provided the same numerical prediction for one and the same ground state {\em observable}, demonstrating that our numerical treatment of ground state properties yields gauge-invariant results. We subsequently turn our attention to computing light and matter subsystem properties in the ground state. These are generally $\alpha$-dependent, but this does {\em not} constitute gauge non-invariance because, as we have discussed in detail in Sec.~\ref{relatphot}, light and matter are quite clearly defined in terms of {\em different observables} in different gauges.

The $\alpha$-dependence of light and matter properties demonstrates the gauge {\em relativity} of these concepts. For every different value of $\alpha$, subsystem predictions are gauge-invariant. For example, the population difference ${\cal P}$ between the first and ground levels of a dipole defined relative to the dipole gauge is represented in the dipole gauge by the operator $\delta'=\ket{e}\bra{e}-\ket{g}\bra{g}$, and in a state ${\cal S}$ represented by density operator $\rho'$ its average value is calculated in the dipole gauge as $\langle{\cal P}\rangle_{\cal S}={\rm tr}(\rho' \delta')$. In the Coulomb gauge the same state and observable are represented by the operators $\rho = R^\dagger \rho' R$ and $\delta = R^\dagger \delta' R$ respectively, and the average is calculated in the Coulomb gauge as $\langle{\cal P}\rangle_{\cal S}={\rm tr}(\rho \delta)$.

\begin{figure}[H]
\raggedright a)\\
\centering
\hspace*{-2.4mm}
\includegraphics[scale=0.7]{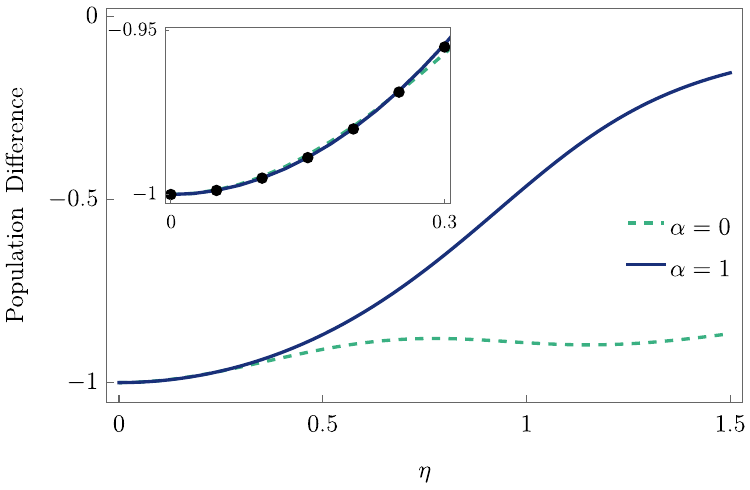}
\vspace*{-0.3cm}

\raggedright b)\\
\centering
\hspace*{-3mm}
\includegraphics[scale=0.7]{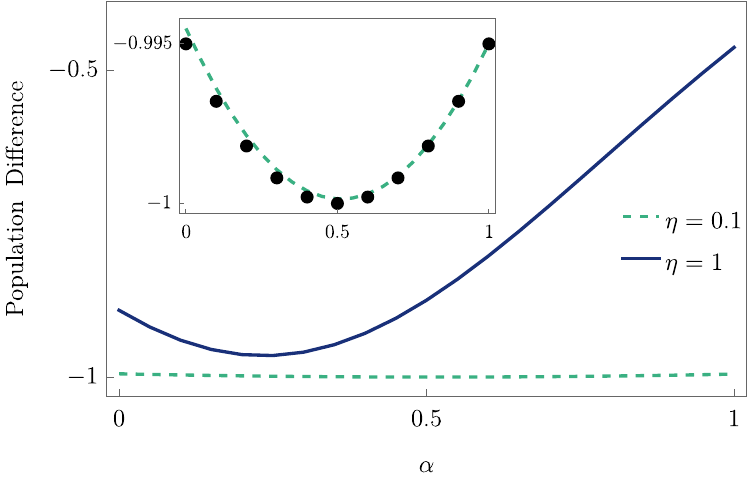}
\vspace*{-0.3cm}
\caption{The population difference $2p_\alpha-1$ between the first excited and ground levels of a highly anharmonic double-well dipole, in the ground state of the dipole-mode composite. We have assumed resonance $\omega = \omega_m$. \textbf{a)} The difference is plotted as a function of $\eta$, for $\alpha=0$ and $\alpha=1$. The inset shows the same curves over the smaller coupling range up to $\eta=0.3$, and the black dotted curve is the weak-coupling two-level dipole prediction of Eq.~(\ref{p}). \textbf{b)} The difference is plotted as a function of $\alpha$ for $\eta=0.1$ (ultrastrong-coupling) and $\eta=1$ (deep-strong coupling). The inset shows the $\eta=0.1$ curve over a smaller range of vertical axis values, and the black dotted curve is the weak-coupling two-level dipole prediction of Eq.~(\ref{p}).}
\label{f1}
\vspace*{-0.3cm}
\end{figure}

The ground state average population difference of a dipole defined relative to the gauge $\alpha$ is shown in Fig.~\ref{f1}. We assume resonance $\omega=\omega_m$ where $\omega_m$ is the transition energy between the lowest two dipole levels. For sufficiently weak coupling the prediction is essentially gauge non-relativistic. In particular the predictions of the second order theory restricted to only the lowest two dipole levels match the numerically calculated predictions for the full Hamiltonian $H_\alpha$, which retain enough levels for convergence and are also non-perturbative. 

\begin{figure}[H]
\raggedright a)\\
\centering
\hspace*{-2.4mm}
\includegraphics[scale=0.805]{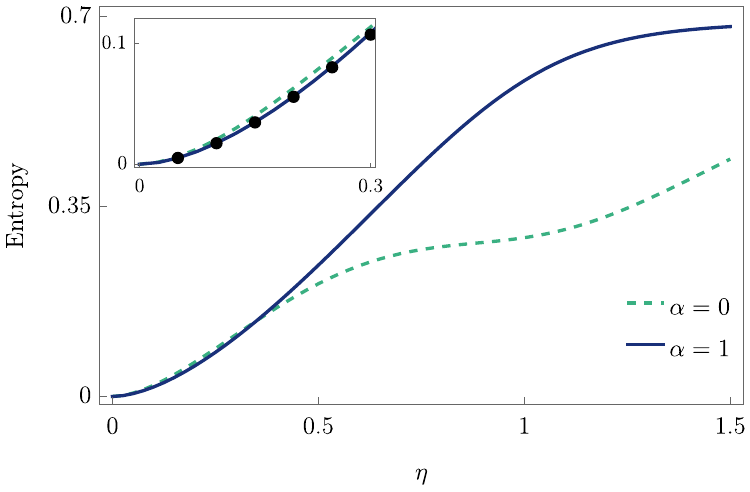}
\vspace*{-0.3cm}

\raggedright b)\\
\centering
\hspace*{-3mm}
\includegraphics[scale=0.8]{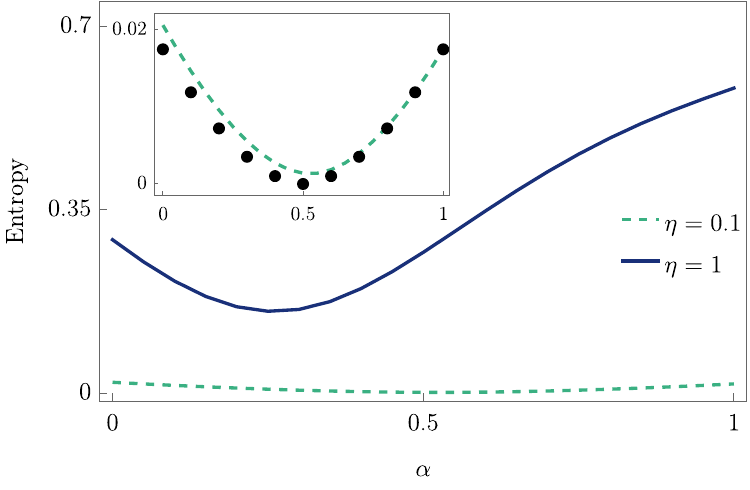}
\vspace*{-0.3cm}
\caption{The entropy of entanglement in the ground state of the dipole-mode composite. We have assumed resonance $\omega = \omega_m$. \textbf{a)} The entropy is plotted as a function of $\eta$, for $\alpha=0$ and $\alpha=1$. The inset shows the same curves over the smaller coupling range up to $\eta=0.3$, and the black dotted curve is the weak-coupling two-level dipole prediction of Eq.~(\ref{p}). \textbf{b)} The entropy is plotted as a function of $\alpha$ for $\eta=0.1$ (ultrastrong-coupling) and $\eta=1$ (deep-strong coupling). The inset shows the $\eta=0.1$ curve over a smaller range of vertical axis values, and the black dotted curve is the weak-coupling two-level dipole prediction of Eq.~(\ref{p}).}
\label{f2}
\vspace*{-0.3cm}
\end{figure}

The population of the excited state drops significantly for intermediate values of $\alpha$ near to $\alpha_{\rm JC}$ wherein the reduced dipole state is closer to the unperturbed ground state (Fig.~\ref{f1} b). For this reason the entropy of entanglement exhibits similar behaviour (Fig.~\ref{f2}). For larger couplings the effect of the ${\bf A}^2$-term becomes more important and when absorbed into a redefinition of the photonic modes it renormalises the mode frequency. This in turn shifts the value of $\alpha \sim \alpha_{\rm JC}$ for which entanglement is minimised, towards the Coulomb gauge value. For all $\alpha$ the absolute value of entanglement becomes larger as the coupling increases and contributions of higher dipole levels become increasingly significant as $\alpha$ approaches zero.

For sufficiently large couplings the entropies of entanglement are found to be significantly different for different $\alpha$ (Fig.~\ref{f2}). The entanglement between the bare dipole and accompanying mode tends to be larger than the entanglement between the corresponding subsystems defined relative to the Coulomb gauge. The near-zone electric field is essentially unavailable for entanglement with the electrostatically dressed dipole, because it is nearly completely cancelled by the static field ${\bf E}_{\rm L}$ that is subtracted from ${\bf E}$ to yield (minus) the Coulomb gauge photonic momentum $-{\bf \Pi}={\bf E}_{\rm T}$.  
 
It is noteworthy that, when focussing on the multipolar and Coulomb gauges at resonance, gauge relativity of entanglement becomes apparent within the full model that retains all material levels, but it is obscured by a two-level dipole truncation. Similarly, gauge-invariance is preserved within the full model, but it is lost within truncated models \cite{stokes_gauge_2019,de_bernardis_breakdown_2018,stokes_gauge_2020,di_stefano_resolution_2019,roth_optimal_2019,li_electromagnetic_2020,ashida_cavity_2021,arwas_metrics_2023}. This example therefore serves to clearly demonstrate the distinction between gauge relativity and gauge non-invariance.


\subsection{Two dipoles and a single mode}

Providing a two-dipole extension of the above model is straightforward. We note again that in the dipole gauge the boundaries are described entirely via the use of appropriate mode functions and there are no direct inter-dipole interactions. Restricting ones attention to a single resonant mode in this gauge preserves the underlying physical structure of the corresponding many-mode theory, which again could in principle be recovered by adding modes one-by-one. The dipoles are labelled by an index $i=1,2$ and ${\bf d}_i$ is the position of the dynamical charge within the $i$'th dipole relative to the centre ${\bf R}_i$, where ${\bf R}_1={\bf 0}$ and ${\bf R}_2={\bf R}$. We assume that the second dipole also resides at a field maximum, ${\bf f}({\bf R})={\bf e}$, and we again suppose that ${\bf d}_i\cdot {\bf e}=d_i$.

The single-mode dipole-gauge model reads
\begin{align}
H'=H_{m1}+H_{m2}+H_{\rm ph}+V'_1+V'_2
\end{align}
where the bare and interaction Hamiltonians are defined as in Eqs.~(\ref{smbhm})-(\ref{smint}) for each subsystem. The $\alpha$-gauge Hamiltonian can be defined by $H_\alpha = R_{1\alpha} H' R_{1\alpha}^\dagger$ where 
\begin{align}
R_{\alpha\alpha'} = \exp\left[i\sum_i (\alpha-\alpha') {\bf d}_i\cdot {\bf A}_{\rm T}\right],
\end{align}
in which ${\bf A}_{\rm T}={\bf A}_{\rm T}({\bf 0})={\bf A}_{\rm T}({\bf R})$. The Hamiltonian $H_\alpha$ includes a direct dipolar interaction term
\begin{align}
V^{\rm inter}_\alpha = -{2\omega \eta^2 \over d_{eg}^2} (1-\alpha^2)d_1 d_2,\label{over}
\end{align}
which for $\alpha=0$ corresponds to the overlap of the Coulomb fields of the dipoles within the toy model. In the gauge $\alpha$ this contribution is found to be weighted by $1-\alpha^2$.

\begin{figure}[t]
\centering
\includegraphics[scale=0.805]{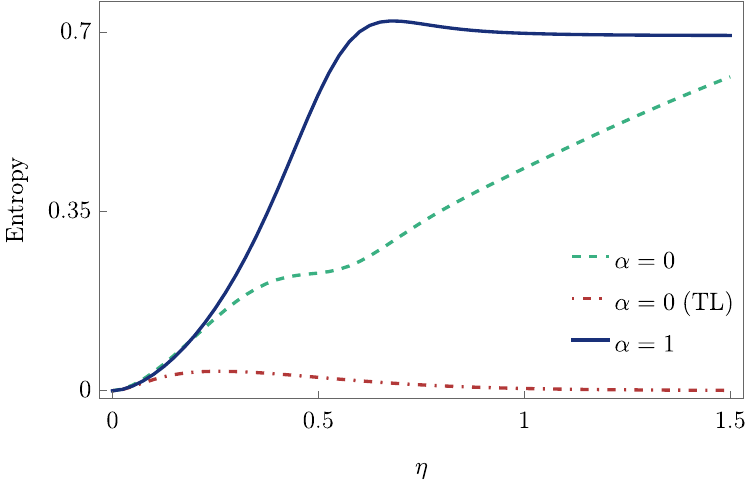}
\caption{The entropy of entanglement of the two dipole reduced density operator (which equals that of the mode's reduced density operator) is plotted with $\eta$ for $\alpha=0$ (green dashed) and $\alpha=1$ (blue solid). The dashed red line shows the $\alpha=0$ curve when the model is truncated to the first two levels of each dipole, showing that the increase in entropy for $\alpha=0$ can be attributed to light-matter coupling to higher material levels.}
\label{fse2}
\end{figure}

We again consider the example of anharmonic double-well dipoles, which we assume are identical and with a first transition energy resonant with the mode. The ground state entropy of entanglement between the mode and dipoles is shown in Fig.~\ref{fse2}. It is essentially gauge non-relativistic for sufficiently small coupling strengths, and increases more rapidly with coupling strength in the dipole gauge than in the Coulomb gauge, mirroring the single-dipole case. It reaches a steady value in the dipole gauge in the deep-strong coupling regime.

\begin{figure}[t]
\centering
\includegraphics[scale=0.805]{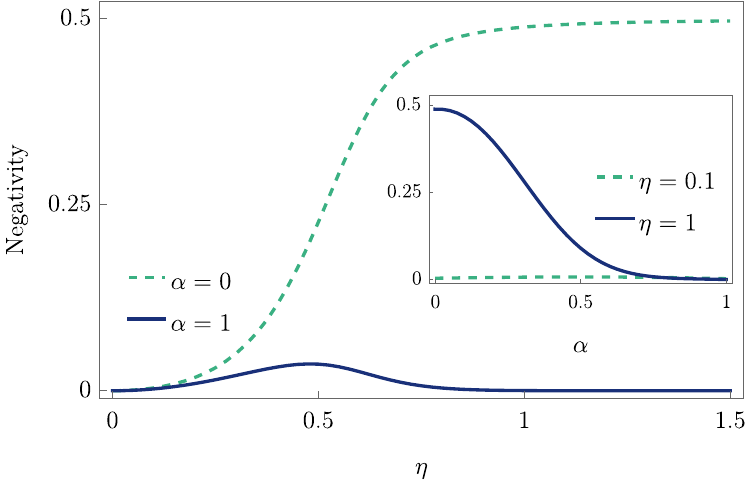}
\caption{The negativity of dipole 1 is plotted with $\eta$ for $\alpha=0$ and $\alpha=1$. We have assumed resonance $\omega=\omega_m$. The inset shows the negativity as a function of $\alpha$. For $\eta=1$ it is maximised in the Coulomb gauge while for $\eta=0.1$ it is maximised near $\alpha=\alpha_{JC}$, which is also near to the minimum of the entropy of entanglement between the mode and dipole.}
\label{fneg2}
\end{figure}

\begin{figure}[H]
\centering
\includegraphics[scale=0.805]{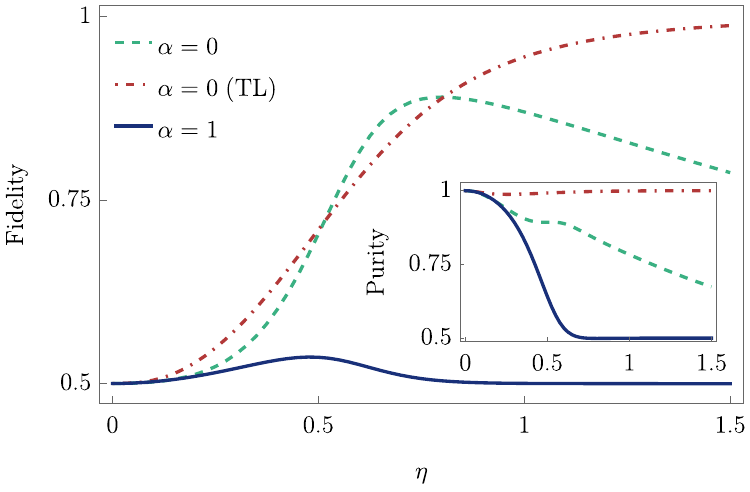}
\caption{The fidelity of the material state in the Bell state $\ket{\psi}$ is plotted with $\eta$. The inset shows the purity of the material state. The red dashed curves are obtained when the $\alpha=0$ theory is truncated to the first two levels of each dipole and are very close to those obtained when, in addition to truncation, one includes only the direct inter-dipole coupling. Within the $\alpha=0$ truncated theory, the material state is essentially $\ket{\psi}$ for large $\eta$, but higher levels become increasingly important for larger $\eta$ and become entangled with the mode (Fig.~\ref{fse2}). This reduces the purity and overlap with $\ket{\psi}$. In the dipole gauge only the lowest two levels of each dipole are important over the shown range of $\eta$, and the curves display behaviour concurrent with the negativity shown in Fig.~\ref{fneg2}.}
\label{fp}
\end{figure}

To determine the inter-dipole entanglement we calculate the negativity associated with dipole 1 defined as
\begin{align}
{\cal N}_{\alpha m} = {1\over 2}\left[\|\rho^{T_1}_{\alpha m}\|-1\right]
\end{align}
where $\|X\|:={\rm tr}(\sqrt{X^\dagger X})$ and ${\rho^{T_1}_{\alpha m}}$ is the partial transpose with respect to dipole $1$, of the reduced dipole density operator $\rho_{\alpha m}$. The negativity is an entanglement monotone and is shown for the Coulomb and dipole gauges in Fig.~\ref{fneg2}. In the dipole gauge it is found to exhibit the same behaviour as the fidelity of the two-dipole reduced state in the maximally entangled Bell state represented by $\ket{\psi}:=(\ket{e,e}+\ket{g,g})/\sqrt{2}$ (Fig.~\ref{fp}). For large $\eta$ the entanglement vanishes as the reduced state approximates the classical mixture $(\ket{\psi}\bra{\psi}+\ket{\phi}\bra{\phi})/2$ where $\ket{\phi}:=(\ket{e,g}+\ket{g,e})/\sqrt{2}$. This state can equivalently be written $(\ket{+,+}\bra{+,+}+\ket{-,-}\bra{-,-})/2$ where $\ket{\pm}=(\ket{e}\pm\ket{g})/\sqrt{2}$, from which it is clear that it is a symmetric mixture of two separable states. 

 In the Coulomb gauge the inter-dipole entanglement as quantified by the negativity is much larger and reaches a steady value within the deep-strong coupling regime. As shown in Fig.~\ref{fp}, much of this entanglement can be attributed to the direct coupling, which moreover, is largely restricted to the lowest two levels of each dipole. This is because $V^{\rm inter}_\alpha$ in Eq.~(\ref{over}) depends only on the dipole moments and so does not scale with dipolar transition energies. Within the two level truncation for each dipole the reduced state for large $\eta$ is essentially the pure Bell state $\ket{\psi}$. This arises almost entirely from the direct dipole-dipole interaction, while the entropy of entanglement associated with the mode vanishes (Fig.~\ref{fse2}). Higher dipole levels become increasingly significant as $\eta$ increases however, and when enough levels for convergence (and gauge-invariance) are included the fidelity of the reduced state in the state $\ket{\psi}$ decreases, as does the purity. The state becomes increasingly mixed and the dipole-mode entanglement increases, albeit much more slowly than in the dipole gauge (Fig.~\ref{fse2}).


\section{Conclusions}\label{conc}

We have studied the gauge relativity of the quantum subsystems called atoms and photons. Different physical definitions of these subsystems, each using different physical observables, are provided by each different gauge. We have specified these observables in terms of the primitive manifestly gauge-invariant and local fields $\rho,{\bf J},{\bf E},$ and ${\bf B}$, finding that different definitions of the atomic subsystem differ in their degree of spatial localisation. The reason can be traced back to Gauss' law, which implies that some generally non-local part, ${\bf P}$, of the local electric field ${\bf E}$, must be a material quantum subsystem observable, but the definition of ${\bf P}$ differs between different gauges.

We have termed the regime in which the gauge relativity of light and matter subsystems can be ignored, {\em gauge non-relativistic}. Beyond this regime, we have discussed the significance of the different definitions of the quantum subsystems for entanglement generation between two atoms ($1$ and $2$). A mediating photonic field is necessary to produce entanglement only if the corresponding atoms are localised to the extent that their material fields ${\bf P}_i,~i=1,2$ do not overlap. We have similarly noted that between such atoms the mediating field propagates causally, yielding purely causal inter-atomic dependencies. However, this implies immediate nonzero excitation of the atom in its local ground state accompanied by the photonic vacuum.

We have studied simple models of a single mode interacting with one and two dipoles, in an arbitrary gauge controlled by a parameter $\alpha$. The choice $\alpha=1$ defines  bare point dipoles each with a material electric field ${\bf P}$ fully localised at the dipole's position. As a result, at every other point in space the source part of the photonic field ${\bf \Pi}$, is the well-known dipolar local electric source-field. The choice $\alpha=0$ yields dipoles fully dressed by their electrostatic fields and the corresponding source part of ${\bf \Pi}$ includes electrostatic contributions that are additional to the local source field of a bare dipole. At the level of individual photonic modes these contributions arise from terms that are singular at zero mode frequency. Such terms occur whenever $\alpha\neq 1$ but can be straightforwardly separated off revealing that they are weighted by $1-\alpha$ in the gauge $\alpha$. We have computed the entropies of entanglement in simple models of a single mode interacting with one and two dipoles. Having found that both light-matter and interatomic entanglement are in general strongly gauge-relative, we have discussed the different underlying physical mechanisms leading to entanglement between different physical subsystems.

\section{Acknowledgments}
We thank Erik Aurell and Ingemar Bengtsson for reading an earlier draft of our manuscript. AS  acknowledges financial support from Newcastle University under a NUAcT fellowship. AN acknowledges discussions with Jonas Larson and financial support from the University of Manchester-Stockholm University Joint Research Fund.

%
%
%
%
%

\end{document}